\documentclass[twocolumn,superscriptaddress,aps,prl,showpacs,preprintnumbers,amsmath,amssymb,floatfix]{revtex4-1}

\usepackage[latin9]{inputenc}
\usepackage{color}
\usepackage{dcolumn}
\usepackage{bm}
\usepackage{bbm}
\usepackage{braket}
\usepackage{mathrsfs}
\usepackage{amstext}
\usepackage{amsmath}
\usepackage{dsfont}
\usepackage{amssymb}
\usepackage{graphicx}
\usepackage{esint}
\usepackage[unicode=true,
 bookmarks=false,
 breaklinks=false,pdfborder={0 0 1},backref=false,colorlinks=true]
 {hyperref}

\makeatletter
\@ifundefined{textcolor}{}
{%
 \definecolor{BLACK}{gray}{0}
 \definecolor{WHITE}{gray}{1}
 \definecolor{RED}{rgb}{1,0,0}
 \definecolor{GREEN}{rgb}{0,1,0}
 \definecolor{BLUE}{rgb}{0,0,1}
 \definecolor{CYAN}{cmyk}{1,0,0,0}
 \definecolor{MAGENTA}{cmyk}{0,1,0,0}
 \definecolor{YELLOW}{cmyk}{0,0,1,0}
}

\newcommand{\eref}[1]{Eq.\,\eqref{#1}}
\newcommand{\mc}[1]{\mathcal{#1}}

\makeatother

\begin{document}

\title{Nearly-linear light cones in long-range interacting quantum systems}

\author{Michael Foss-Feig}
\affiliation{Joint Quantum Institute, NIST/University of Maryland, College Park,
Maryland 20742, USA}

\author{Zhe-Xuan Gong}
\affiliation{Joint Quantum Institute, NIST/University of Maryland, College Park,
Maryland 20742, USA}

\author{Charles W. Clark}
\affiliation{Joint Quantum Institute, NIST/University of Maryland, College Park,
Maryland 20742, USA}

\author{Alexey V. Gorshkov}
\affiliation{Joint Quantum Institute, NIST/University of Maryland, College Park,
Maryland 20742, USA}

\date{\today}
\begin{abstract}

In non-relativistic quantum theories with short-range Hamiltonians, a velocity $v$ can be chosen such that the influence of any local perturbation is approximately confined to within a distance $r$ until a time $t \sim r/v$, thereby defining a linear light cone and giving rise to an emergent notion of locality.  In systems with power-law ($1/r^{\alpha}$) interactions, when $\alpha$ exceeds the dimension $D$, an analogous bound confines influences to within a distance $r$ only until a time $t\sim(\alpha/v)\log r$, suggesting that the velocity, as calculated from the slope of the light cone, may grow exponentially in time.  We rule out this possibility; light cones of power-law interacting systems are algebraic for $\alpha>2D$, becoming linear as $\alpha\rightarrow\infty$.  Our results impose strong new constraints on the growth of correlations and the production of entangled states in a variety of rapidly emerging, long-range interacting atomic, molecular, and optical systems.

\end{abstract}

\pacs{03.67.-a, 05.70.Ln, 75.10.Pq}

\maketitle

\begin{figure}[t!]
\includegraphics[width=1.0\columnwidth]{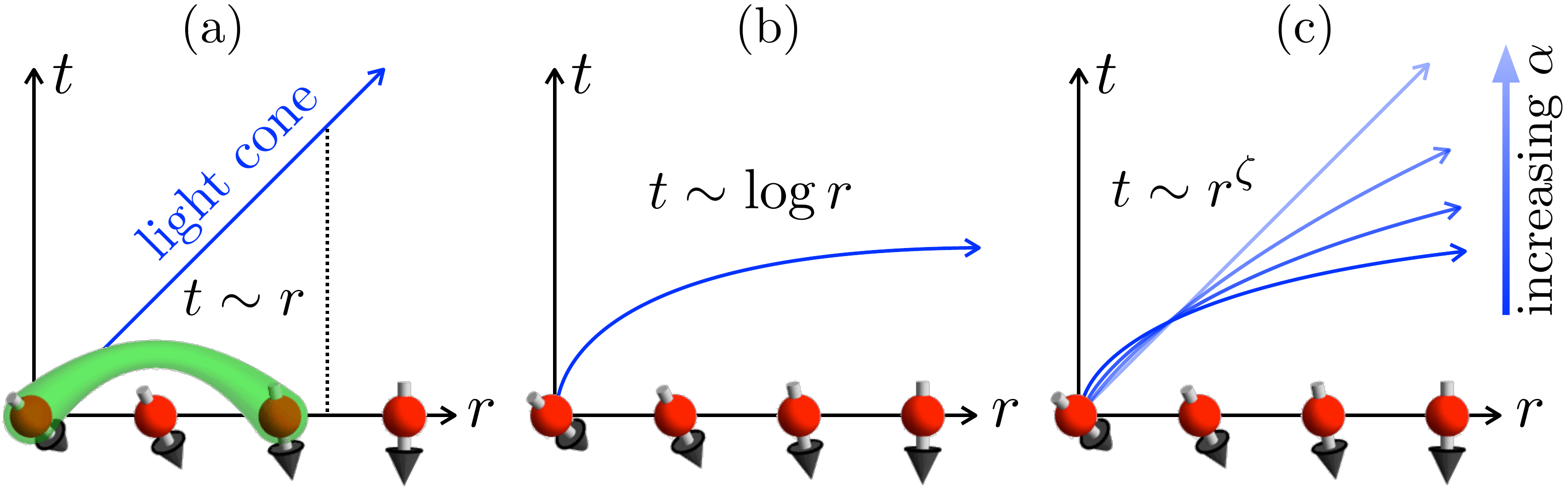}
\caption{(color online). (a) In a short-range interacting system, perturbing a single spin at $t=r=0$ can only influence another spin (green connection) if it falls within a causal region bounded by a linear light cone ($t\sim r$) \cite{lieb72}.  (b) Existing bounds for power-law interacting systems \cite{Hastings05,gong14} result in a logarithmic light cone ($t\sim \log r$) at large distances and times, and thus the maximum velocity grows exponentially in time.  (c) We show that light cones of power-law interacting systems are necessarily polynomial, becoming increasingly linear for shorter-range interactions.}
\label{fig:fig1}
\end{figure}

Though non-relativistic quantum theories are not explicitly causal, Lieb and Robinson \cite{lieb72} proved that an effective speed limit emerges dynamically in systems with short-ranged interactions, thereby extending the notion of causality into the fields of condensed matter physics, quantum chemistry, and quantum information science.  Specifically, they proved that when interactions have a finite range or decay exponentially in space, the influence of a local perturbation decays exponentially outside of a space-time region bounded by the line $t=r/v$, which therefore plays the role of a light cone [Fig.\,\ref{fig:fig1}(a)].  However, many of the systems to which non-relativistic quantum theory is routinely applied---ranging from frustrated magnets and spin glasses \cite{RevModPhys.58.801,PhysRev.96.99} to numerous atomic, molecular, and optical systems \cite{saffman10,islam13,gopalakrishnan11,aikawa12,yan13}---possess power-law interactions, and hence do not satisfy the criteria set forth by Lieb and Robinson.  Many questions about the fate of causality in such systems lack complete answers:  Can information be transmitted with an arbitrarily large velocity \cite{eisert13}, and if so, how quickly (in space or time) does that velocity grow? Under what circumstances does a causal region exist, and when it does, what does it look like \cite{eisert13,hauke13,juenemann13,gong14,Richerme14,Jurcevic14}?   The answers to these questions have far reaching consequences, for example imposing speed limits on quantum-state transfer  \cite{bose07} and on thermalization rates in many-body quantum systems \cite{polkovnikov11}, determining the strength and range of correlations in equilibrium \cite{Hastings05}, and constraining the complexity of simulating quantum dynamics with classical computers \cite{PhysRevLett.100.070502}.

The results of Lieb and Robinson were first generalized to power-law ($1/r^{\alpha}$) interacting systems by Hastings and Koma \cite{Hastings05}, with the following picture emerging.  For $\alpha>D$, the influence of a local perturbation is bounded by a function $\propto e^{vt}/r^{\alpha}$, and while a light cone can still be defined as the boundary outside of which this function falls below some threshold value, yielding $t\sim\log r$, that boundary is \emph{logarithmic} rather than linear [Fig.\,\ref{fig:fig1}(b)].  Improvements upon these results exist, revealing, e.g., that the light-cone remains linear at intermediate distance scales \cite{gong14}, but all existing bounds consistently predict an asymptotically logarithmic light cone.  An immediate and striking consequence is that the maximum group velocity, defined by the slope of the light cone, grows exponentially with time, thus suggesting that the aforementioned processes --- thermalization, entanglement growth after a quench, etc. --- may in principle be sped up \emph{exponentially} by the presence of long-range interactions.  In this manuscript, we show that this scenario is not possible.  While light cones can potentially be sub-linear for any finite $\alpha$, thus allowing a velocity that grows with time, for $\alpha>2D$ they remain bounded by a polynomial $t\sim r^{\zeta}$, and $\zeta\leq 1$ approaches unity for increasing $\alpha$ [Fig.\,\ref{fig:ball}(c)].

%  On the one hand, the generalized Lieb-Robinson result suggests that the causal region may be bounded by a function $t\sim\alpha\log r$, with signals decaying polynomially outside of a causal region that grows exponentially in time.  On the other hand, an extensive collection of numerical calculations for specific interacting spin models suggests that the linear light-cone is only slightly modified by power-law interactions, and certainly show no indication of an exponential speedup of the light-cone edge.

\emph{Model and formalism}.---We assume a generic spin model with time-independent Hamiltonian \footnote{The results presented are significantly more general than \eref{eq:hamiltonian} suggests, and can easily be generalized to fermionic models, or Hamiltonians with arbitrary single-particle terms or time dependence.}
\begin{equation}
\label{eq:hamiltonian}
H=\frac{1}{2}\sum_{\mu,y,z}J_{\mu}(y,z)V_{y\mu} V_{z\mu},
\end{equation}
where $V_{y\mu}$ is a spin operator on site $y$ with $\lVert V_{y\mu}\rVert=1$ (where $\lVert O\rVert$ denote the operator norm of an operator $O$, which is the magnitude of its eigenvalue with largest absolute value).  The coupling constants must satisfy $\sum_{\mu}J_{\mu}(y,z)\equiv J(y,z)\leq J/d(y,z)^{\alpha}$ for $y\neq z$, with  $d(y,z)$ the distance between lattice sites $y$ and $z$, and $J_{\mu}(y,y)=0$.  Our goal is to bound the size of an unequal-time commutator of two unity-norm operators $A$ and $B$ initially residing on sites $i$ and $j$, respectively,
\begin{align}
\label{eq:utc}
C_r(t)=\lVert [A(t),B]\rVert\leq \mathscr{C}_r(t),
\end{align}
where $r=d(i,j)$.  Since spin operators on different sites commute, $C_r(t)$ captures the extent to which an operator $A$ has ``spread'' onto the lattice site $j$ during the time evolution.  As a result, it bounds numerous experimentally measurable quantities, for example connected correlation functions after a quantum quench \cite{bravyi06,gong14,Richerme14,Jurcevic14}.  In general, a light cone can be defined by setting $\mathscr{C}_r(t)$ equal to a constant and solving for $t$ as a function of $r$.  A natural way to parametrize the shape of the light cone is to ask whether it can be bounded by the curve $r=t^{\beta}$ (with $\beta\geq0$) in the large $t$ limit, which is true whenever $\lim_{t\rightarrow\infty}\mathscr{C}_{t^{\beta}}(t)=0$.  Defining $1/\zeta$ to be the smallest value of $\beta$ for which this limit vanishes, we can say that $t\sim r^{\zeta}$ is the tightest possible polynomial light cone.  The original work by Lieb and Robinson proved that $\zeta=1$ when interactions are finite-ranged or exponentially decaying.  However, the generalization of their results to power-law interacting Hamiltonians \cite{Hastings05} yields $\mathscr{C}_r(t)\sim e^{vt}/r^{\alpha}$,  and thus $\lim_{t\rightarrow\infty}\mathscr{C}_{t^{\beta}}(t)$ never vanishes for finite $\beta$.  Though Ref. \cite{gong14} demonstrated that a linear light cone can still persist at intermediate distance scales, the true asymptotic shape of the light cone was nevertheless logarithmic.  Thus the consensus of all previously available bounds is that $\zeta\rightarrow 0$, and the light cone is not bounded by a polynomial.

%\emph{Origin of the logarithmic light-cone}.--- In order to understand our results, it is helpful to first consider why the Lieb-Robinson bound for power-law interacting systems gives rise to a logarithmic light-cone.  Once interactions become long-ranged, the it is possible to have high-order contributions where that remain inside the nearest neighbor light-cone, and then reach the final site with a single matrix element $\sim r^{-\alpha}$

\emph{Strategy}.---To prove the existence of a polynomial light cone, we begin by breaking $H$ into a short-range and a long-range contribution, $H=H^{\rm sr}+H^{\rm lr}$, separated by a cutoff length scale $\chi$.  Defining $J^{\rm sr[lr]}_{\mu}(y,z)=J_{\mu}(y,z)$ if $d(y,z)\leq\chi~ [>\chi ]$ and $0$ otherwise,  we can write
\begin{align}
H^{\rm sr[lr]}=\sum_{\mu,y,z}J_{\mu}^{\rm sr[lr]}(y,z)V_{y\mu} V_{z\mu}.
\end{align}
We then move to the interaction picture of $H^{\rm sr}$, where
\begin{equation}
\label{eq:Crt2}
C_r(t)=\lVert [\mc{U}^{\dagger}(t)\mc{A}(t)\,\mc{U}(t),B]\rVert.
\end{equation}
Here $\mc{A}(t)=\exp(i t H^{\rm sr})A\exp(-i t H^{\rm sr})$ [and all other script operators except $\mc{U}(t)$] is evolving under the influence of $H^{\rm sr}$, and the interaction-picture time evolution operator $\mc{U}(t)$ is a time-ordered exponential
\begin{equation}
\mathcal{U}(t)=T_{\tau} \exp\left(-i\int_{0}^{t} \!\! d\tau\,\mathcal{H}^{\rm lr}(\tau) \right),
\end{equation}
where
\begin{align}
\mathcal{H}^{\rm lr}(\tau)=\frac{1}{2}\sum_{\mu,y,z} J_{\mu}^{\rm lr}(y,z)\mc{V}_{y\mu}(\tau)\mc{V}_{z\mu}(\tau)\equiv\sum_{y,z} \mc{W}_{yz}(\tau).
\end{align}

%Note that in \eref{eq:Crt2} we are allowing for different times, $s$ and $t$, for evolution under the short-range Hamiltonian and evolution due to the interaction-picture time evolution operator, respectively.  This subtle distinction is conceptually important, because we will compute the equations of motion for $\mc{U}(t)$ (in order to constrain the dynamics it induces) while holding fixed the time-dependence of $\mc{A}(s)$.  At the end of the calculation, we can recover $C_r(t)$ simply by setting $s=t$.

\begin{figure}[t!]
\includegraphics[width=1.03\columnwidth]{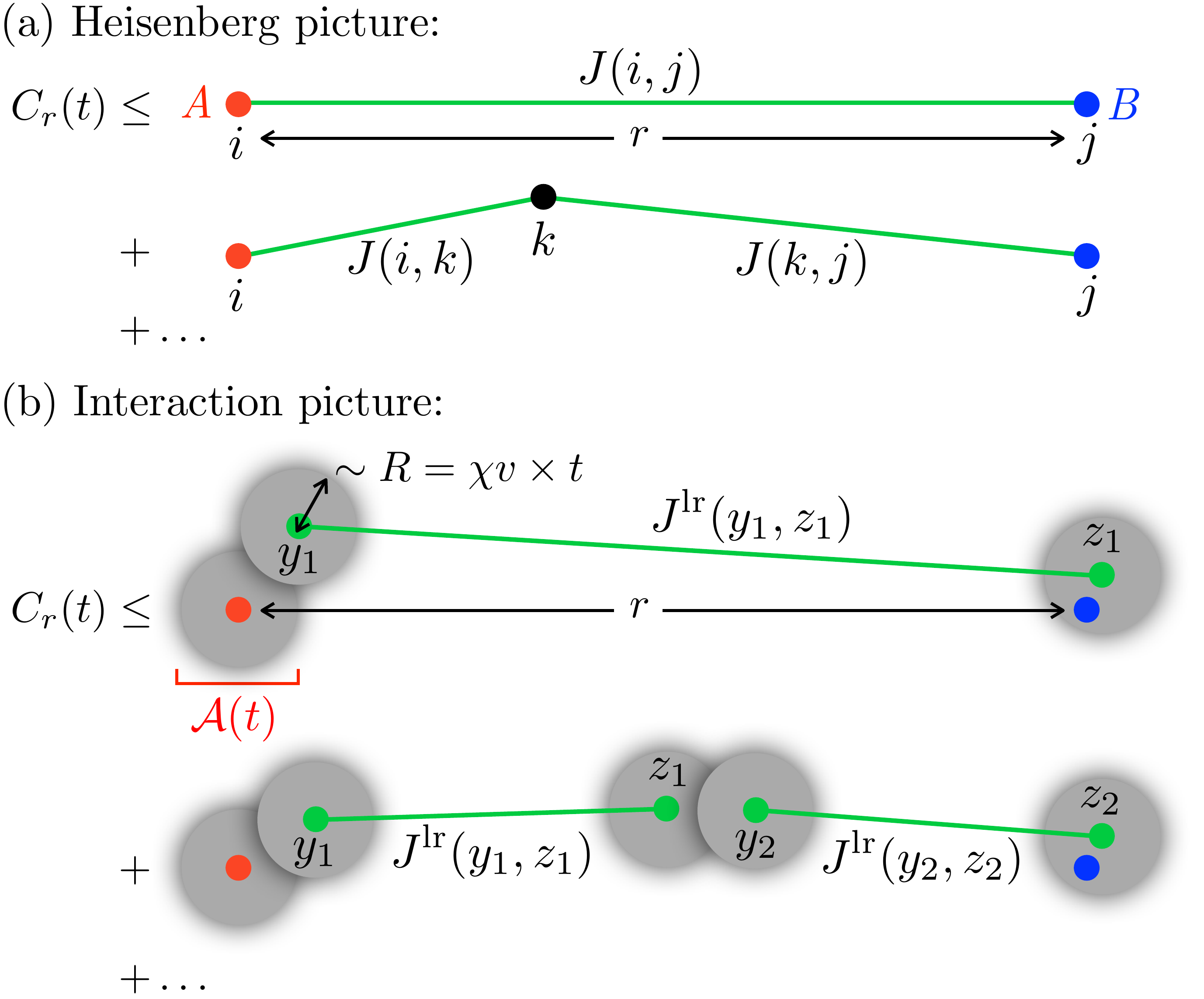}
\caption{(color online) Schematic illustration of a Lieb Robinson-type bound. (a) \emph{Heisenberg picture}. The time evolution of an operator $A$ is bounded by a series in which repeated applications of the Hamiltonian connect site $i$ to site $j$. (b) \emph{Interaction picture}. A similar series can be used to bound the dynamics induced by the interaction-picture time-evolution operator, but now the operator $A$ and the interaction terms in the Hamiltonian are spread out over a light-cone radius of the short-range Hamiltonian.}
\label{fig:schematic}
\end{figure}

The plan is now to treat the short-range physics, responsible for the time-dependence of interaction picture operators $\mc{A}(t)$ and $\mc{W}_{yz}(\tau)$, and the long-range physics, captured by the remaining interaction-picture time evolution operator $\mc{U}(t)$, with two independent bounds.  The basic physical picture to have in mind is shown in Fig.\,\ref{fig:schematic}.  The original Lieb-Robinson approach is to work in the Heisenberg-picture, expressing $\mathscr{C}_r(t)$ as series of terms connecting sites $i$ and $j$ by repeated applications of $H$ \cite{lieb72,Hastings05,nachtergaele10, gong14} [Fig.\,\ref{fig:schematic}(a)].  We will instead bound the dynamics induced by $\mc{U}(t)$ by a series of terms connecting sites $i$ and $j$ by repeated applications of $\mc{H}^{\rm lr}(\tau)$ [Fig.\,\ref{fig:schematic}(b)].  Though $\mc{H}^{\rm lr}(\tau)$ is not a sum of local operators, the $\mc{V}_y(\tau)$ which comprise it are still approximately contained within a ball of radius $R= \chi v \times t $ [gray shaded disks in Fig.\,\ref{fig:schematic}(b)], which is the light cone of the short-range Hamiltonian.  Here $v$ would be the Lieb-Robinson velocity for a nearest-neighbor Hamiltonian with coupling strength $J$, and must be multiplied by $\chi$ to account for the longest-range terms in $H^{\rm sr}$.

Our approach is motivated by the following observation: If we assume the existence of a logarithmic light cone, we can choose the cutoff $\chi$ to scale in such a way that $\mathscr{C}_r(t)$ does not grow exponentially in time, which contradicts the assumption.  To see this, we first note that the existence of a logarithmic light cone allows us to choose $\chi$ to scale with \emph{any} power of $t$ while satisfying the following inequality along the light-cone boundary (at sufficiently long times),
\begin{equation}
\label{eq:separation}
R=\chi v\times t\ll r\sim e^{vt}.
\end{equation}
Physically, this inequality ensures that the point $r$ falls well outside the short-range light-cone distance $R$, and as a result both the operator $\mc{A}(t)$ and the $\mathcal{V}_y(\tau)$ comprising $\mathcal{H}^{\rm lr}(\tau \leq t)$ appear nearly local when viewed on the length scale $r$.  We therefore expect that the time evolution induced by $\mathcal{U}(t)$  [Fig.\,\ref{fig:schematic}(b)] should be qualitatively similar---up to the possibility of a different velocity---to that induced by $U$ [Fig.\,\ref{fig:schematic}(a)].  The velocity can be estimated by considering the following expansion of $A(t)$,
\begin{align}
A(t)=\mc{A}(t)+i\sum_{yz}\int_{0}^{t}d\tau[\mc{W}_{yz}(\tau),\mc{A}(t)]+\dots.
\label{eq:velocity_approx}
\end{align}
Due to the quasi-locality of interaction-picture operators, a general commutator $[\mc{W}_{yz}(\tau),\mc{A}(t)]$ is exponentially suppressed unless either $y$ or $z$ resides within a distance $2R$ of site $i$.  Ignoring (for now) the exponentially small corrections from outside the short-range light cone, we can restrict the summation to run over $y$ and $z$ such that either $d(i,y)\leq 2R$ or $d(i,z)\leq 2R$, giving
\begin{align}
\label{eq:simple_velocity_estimate}
\lVert\sum_{y,z}\int_{0}^{t}d\tau[\mc{W}_{yz}(\tau),\mc{A}(t)]\rVert\!\lesssim t\times R^{D}\lambda_{\chi},
\end{align}
where $\lambda_{\chi}=\sum_{z}J^{\rm lr}(y,z)\sim  \chi^{D-\alpha}$ [with $J^{\rm sr[lr]}(y,z)=\sum_{\mu}J_{\mu}^{\rm sr[lr]}(y,z)$]. The coefficient of $t$ on the right-hand side of \eref{eq:simple_velocity_estimate} suggests a velocity $v_{\chi}\sim R^{D}\lambda_{\chi}\sim t^{D} \chi^{2D-\alpha}$, which can be made small for large $\chi$ whenever $\alpha>2D$. 

An important achievement of this paper is a proof that the parametrically small velocity $v_{\chi}$ also controls the higher-order contributions from the interaction picture time-evolution operator.  Therefore, in moving from the Heisenberg picture to the interaction picture, we are able (loosely speaking) to make the replacement $\mathscr{C}_r(t)\sim\exp(vt)/r^{\alpha}\rightarrow \exp( v_{\chi}t)/r^{\alpha}$.  By letting $\chi$ grow with $t$ in such a way that $v_{\chi}t$ stays constant in time [which can always be done in a manner consistent with \eref{eq:separation}], the exponential time dependence is suppressed, violating our assumption of a logarithmic light cone.  Indeed, as we will show, a proper scaling of $\chi$ will enable us to change the time dependence from exponential to algebraic, which in turn enables the recovery of a polynomial light-cone.

\emph{Derivation}.---In order to formalize the above picture, we must first take a step back and treat the interaction-picture operators more carefully.  First, we denote the set of points within a radius $R+\ell \chi$ of the point $i$ by $\mathscr{B}_{\ell}(i)$, and the complement of this set by $\overline{\mathscr{B}}_{\ell}(i)$.  Now we can obtain an approximation to $\mc{A}(t)$, supported entirely on $\mathscr{B}_{\ell}(i)$, by integrating over all unitaries on $\overline{\mathscr{B}}_{\ell}(i)$ with respect to the Haar measure \cite{bravyi06,SOM}, $\mathcal{A}(\ell,t)=\int_{\overline{\mathscr{B}}_{\ell}(i)} d\mu(U)U\mc{A}(t)U^{\dagger}$.  Because $\mc{A}(\ell,t)$ tends to $\mc{A}(t)$ as $\ell\rightarrow\infty$, we can rewrite $\mc{A}(t)=\sum_{\ell=0}^{\infty}\mc{A}^{\ell}(t)$, with $\mc{A}^{0}(t)=\mc{A}(0,t)$ and $\mc{A}^{\ell>0}(t)=\mc{A}(\ell,t)-\mc{A}(\ell-1,t)$.  Each operator $\mc{A}^{\ell}(t)$ is supported on $\mathscr{B}_{\ell}(i)$, and is expected to become small for large $\ell$, since both $\mc{A}(\ell,t)$ and $\mc{A}(\ell-1,t)$ are becoming better approximations to $\mc{A}(t)$, and hence must be approaching each other.  Formally, by applying a standard short-range Lieb-Robinson bound to $H^{\rm sr}$, one can show that $\lVert \mc{A}^{\ell}(t)\rVert\leq c e^{-\ell}$, with $c$ a constant \cite{SOM}.  The ability to write $\mc{A}(t)$ as the sum of a sequence of operators with increasing support but exponentially decreasing norm is the mathematical basis for the intuition that interaction-picture operators are quasi-local.  A similar construction enables us to write $\mc{W}_{yz}(\tau)=\sum_{m,n}\mc{W}_{\xi}(\tau)$, where the index $\xi=\left\{y,z,m,n\right\}$ describes the location $y[z]$ and support $m[n]$ of the operators $\mc{V}_{y}^{m}(\tau)[\mc{V}_{z}^{n}(\tau)]$ comprising $\mc{W}_{\xi}(\tau)$.  Once again, the size of these operators decreases exponentially in the radius of their support,
\begin{align}
\lVert \mc{W}_{\xi}(\tau)\rVert\leq c^2 J^{\rm lr}(y,z)e^{-(m+n)}/2,
\end{align}
but algebraically in the separation $d(y,z)$.

Now we would like to constrain the time evolution due to $\mathcal{U}(t)$, which further expands the support of $\mathcal{A}(t)$ in \eref{eq:Crt2}.  As suggested in Fig.\,\ref{fig:schematic}, our bound is comprised of terms in which sites $i$ and $j$ are connected by repeated applications of the interaction-picture Hamiltonian $\mc{H}^{\rm lr}(\tau)$.  Employing a generalization of the techniques originally used by Lieb and Robinson, we obtain \cite{SOM}
\begin{align}
\label{eq:prelim_bound}
C_r(t)\leq \sum_{\ell}\lVert[\mc{A}^{\ell}(t),B]\rVert+4c\sum_{a=1}^{\infty}\frac{t^a}{a!}\mc{J}_a(i,j),
\end{align}
where
\begin{align}
&\mc{J}_{a}(i,j)=4^a\!\!\!\!\sum_{\ell,\xi_1,\dots,\xi_a}\!\!\!\!e^{-\ell}D_{\rm i}(\xi_1)\lVert\mc{W}_{\xi_1}\rVert D(\xi_1,\xi_2)\lVert\mc{W}_{\xi_2}\rVert \times\dots\nonumber\\
\label{eq:J_def}
&\dots\times \lVert\mc{W}_{\xi_{a-1}}\rVert D(\xi_{a-1},\xi_{a}) \lVert\mc{W}_{\xi_a}\rVert D_{\rm f}(\xi_{a}).
\end{align}
Here $D(\xi_1,\xi_{2})$ is unity whenever $\mathscr{B}_{n_1}(z_1)\cap\mathscr{B}_{m_2}(y_2)\neq\varnothing$ and vanishes otherwise, thus constraining the points $z_1$ and $y_{2}$ in the progression $\dots\lVert\mc{W}_{\xi_1}\rVert D(\xi_1,\xi_{2})\lVert \mc{W}_{\xi_{2}}\rVert\dots$ to be near each other, as shown in Fig.\,\ref{fig:ball}(a).  Similarly, $D_{\rm i}(\xi_1)$ is unity when $\mathscr{B}_{\ell}(i)\cap\mathscr{B}_{m_1}(y_1)\neq\varnothing$ and vanishes otherwise, while $D_{\rm f}(\xi_a)$ is unity when $j\in \mathscr{B}_{n_a}(z_a)$ and vanishes otherwise, thus constraining the first interaction $\mc{W}_{\xi_1}$ to originate from near the point $i$, and the final one $\mc{W}_{\xi_a}$ to terminate near the point $j$.

\begin{figure}[!t!]
\includegraphics[width=1.0\columnwidth]{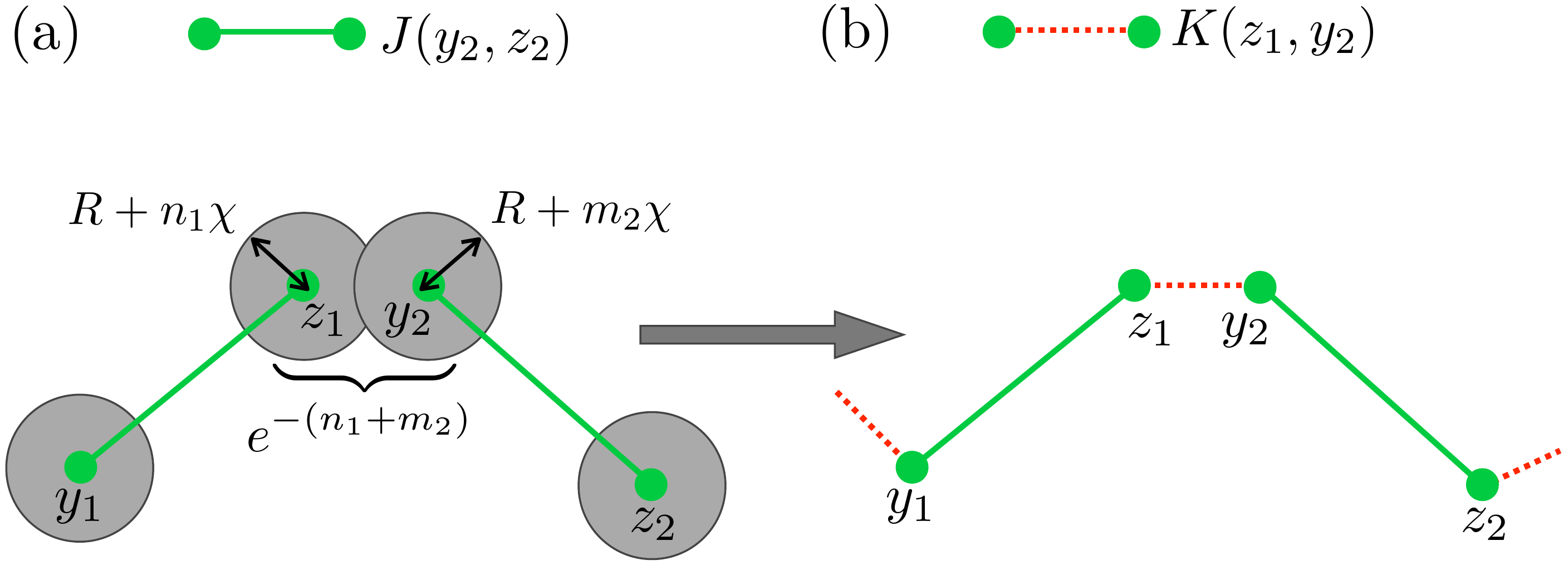}
\caption{(color online). (a) Schematic representation of the term $\dots\lVert\mc{W}_{\xi_1}\rVert D(\xi_1,\xi_{2})\lVert \mc{W}_{\xi_{2}}\rVert\dots$in \eref{eq:J_def}.  Each green line represents a single term in the interaction-picture Hamiltonian, and the operators at the endpoints are supported over a ball or radius $R+m\chi$ (grey disks). (b) In deriving a bound, the additional summations over the sizes of the supports of each operator generate exponentially decaying connections between successive terms [\eref{eq:D_to_K}].}
\label{fig:ball}
\end{figure}

Equation \eqref{eq:J_def} can be simplified by first carrying out the summation over indices $m_1,\dots,m_a$ and $n_1,\dots,n_a$, which were necessary to account for the exponentially decaying contribution to interaction-picture operators outside the short-range light cone.  For example, considering the intersection shown in Fig.\,\ref{fig:ball}(a), one can show that
\begin{align}
\label{eq:D_to_K}
\sum_{n_1,m_2}&\lVert \mc{W}_{\xi_1}\rVert D(\xi_1,\xi_2)\lVert \mc{W}_{\xi_2}\rVert\leq\\
&\kappa^2\frac{c^2 e^{-m_1}}{2}J^{\rm lr}(y_1,z_1)K(z_1,y_2)J^{\rm lr}(y_2,z_2)\frac{c^2 e^{-n_2}}{2},\nonumber
\end{align}
(with $\kappa$ a constant), where $K(z_1,y_2)$ decays exponentially in $d(z_1,y_2)$ \cite{SOM}, directly reflecting the quasi-locality of the interaction-picture operators. Using this inequality repeatedly in \eref{eq:J_def} we obtain
\begin{align}
\mc{J}_{a}(i,j)&\!\leq \kappa^{2}(2\kappa^2 c^2)^{a}\!\!\!\!\sum_{\substack{y_1,\dots, y_a \\ z_1,\dots, z_a}}\!\!\!\!\!K(i,y_1)J^{\rm lr}\!(y_1,z_1)K(z_1,y_2)\times\nonumber\\
\label{eq:JK_def}
\dots\times\! J^{\rm lr}\!&(y_{a-1},z_{a-1})K(z_{a-1},y_{a}) J^{\rm lr}\!(y_a,z_a)K(z_a,j).
\end{align}
Every term in \eref{eq:JK_def} connects sites $i$ and $j$ by repeated applications of $K$'s and $J$'s, which capture, respectively, physics below and above the cutoff length-scale $\chi$ [see Fig.\,\ref{fig:ball}(b)].  The summations over indices $y_1,\dots,y_a$ can then be carried out by bounding the discrete convolution $\sum_{y_2}K(z_1,y_2)J^{\rm lr}(y_2,z_2)\leq (2\kappa\lambda_{\chi})F(z_1,z_2)$ to give
\begin{align}
\mc{J}_{a}(i,j)\!\leq\! 2\kappa^{2}(\!4\kappa^3 c^2\lambda_{\chi})^{a}\!\!\!\!\!\sum_{z_1,\dots, z_a}\!\!\!\!\!F(i,z_1)\!\times\!\dots\!\times\! F(z_a,j).
\label{eq:just_Fs}
\end{align}
Because $K$ decays exponentially while $J$ decays only algebraically, their convolution is dominated [at large $d(z_1,z_2)$] by terms where $y_2$ is much closer to $z_1$ than to $z_2$, and hence $F$ inherits the long-distance algebraic decay of $J^{\rm lr}$ \footnote{As $\alpha$ gets larger, larger separations are required for the algebraic decay to dominate over the exponential decay.  Strictly speaking, this consideration imposes that \eref{eq:final_bound} is only valid whenever $vt>\alpha\log\alpha$. This restriction is irrelevant, since we are ultimately concerned with the asymptotic light-cone shape at large $r$ and $t$, however the $\alpha\rightarrow\infty$ limit could in principle be taken at finite $r$ and $t$ by using the techniques of Ref. \cite{gong14}.},
\begin{align}
F(z_1,z_2)=\left\{\!\!\begin{array}{cc}
     1;&d(z_1,z_2)\leq6R\\
     \left[6R/d(z_1,z_2)\right]^{\alpha}; &d(z_1,z_2)>6R.\\
\end{array}\right.
\end{align}
The remaining summation over indices $z_1,\dots, z_a$ can be carried out (as in Ref. \cite{Hastings05}) by invoking a so-called reproducibility condition, valid for power-law decaying functions.  In particular, we find $\sum_{z_2}F(z_1,z_2)F(z_2,z_3)\leq gR^{D} F(z_1,z_3)$, where $g$ is a constant and the factor of $R^{D}$ enters because $F(z_1,z_2)$ decays algebraically only for $d(z_1,z_2)\gtrsim R$.  Utilizing this condition repeatedly in \eref{eq:just_Fs}, we obtain (for $r>6R$)
\begin{align}
\label{eq:final_J_bound}
\mc{J}_a(i,j)\leq \kappa^2\,(R/r)^{\alpha}\times (v_{\chi})^{a},
\end{align}
where further numerical pre-factors have been absorbed into $\kappa^2$, and $v_{\chi}=\vartheta R^{D}\lambda_{\chi}$ is a cutoff-dependent velocity (with $\vartheta$ a constant) with the scaling predicted by \eref{eq:simple_velocity_estimate}.   Plugging \eref{eq:final_J_bound} into \eref{eq:prelim_bound}, we obtain our final bound
\begin{align}
\label{eq:final_bound}
C_r(t)\leq\mathscr{C}_r(t)\equiv 2c\kappa\!\left(e^{vt-r/\chi}+2\kappa \frac{e^{v_{\chi}t}}{(r/R)^{\alpha}}\right).
\end{align}
The first term is the bound one would obtain for the finite-range Hamiltonian $H^{\rm sr}$.  The second term contains the effect of $H^{\rm lr}$, which leads to a bound similar to that of Ref. \cite{Hastings05}, except with a velocity that is parametrically small in the cutoff $\chi$, and a distance $r$ that is rescaled by the radius $R$ of the short-range light cone.

\emph{Light cone shape}.---Equation \eqref{eq:final_bound} can now be minimized with respect to the cutoff $\chi$, which we accomplish by letting $\chi$ scale with time as a power-law ($\chi\propto t^{\gamma}$), which enforces the scaling $R\sim t^{\gamma+1}$ and $v_{\chi}t\sim t^{(1+D)+\gamma(2D-\alpha)}$.  The exponential time dependence can be suppressed by keeping $v_{\chi}t\sim 1$, which requires $\gamma=(1+D)/(\alpha-2D)$.  Dropping pre factors (since we only care about asymptotics at large $r$ and $t$), we obtain
\begin{equation}
\label{eq:final_scaling_bound}
\mathscr{C}_r(t)\sim\exp[vt-r/t^{\gamma}]+\frac{t^{\alpha(1+\gamma)}}{r^{\alpha}}.
\end{equation}
Thus, as argued earlier, the cutoff can be chosen to scale with time in such a way that the long-range contribution to the bound (scaling in space as $r^{-\alpha}$) has an \emph{algebraic} rather than exponential time dependence.  If we now make the substitution $r=t^{\beta}$, we see that $\lim_{t\rightarrow\infty}\mathscr{C}_{t^{\beta}}(t)$ vanishes whenever $\beta>1/\zeta$, with 
\begin{align}
\label{eq:light_cone_shape}
1/\zeta=1+(1+D)/(\alpha-2D).
\end{align}
Thus the light cone is bounded by a power law $t=r^{\zeta}$ ($0<\zeta<1$) whenever $\alpha>2D$, and becomes increasingly linear ($\zeta\rightarrow 1$) as $\alpha$ grows larger.

As discussed in the introduction, our results impose stringent constraints on the growth of entanglement after a quantum quench.  In addition, our bound implies much stricter constraints on \emph{equilibrium} correlation functions than were previously known \cite{Hastings05}.  In particular, it follows from Eqs.\,\eqref{eq:final_scaling_bound} and \eqref{eq:light_cone_shape} that correlations in the ground state of $H$ decay at long distances as $1/r^{\alpha}$, so long as the spectrum of $H$ remains gapped \cite{Gong_inprep_14} (in fact, when combined with the results of Ref. \cite{gong14}, the bound derived here could be used to show that ground-state correlation functions exhibit a hybrid exponential-followed-by-algebraic decay, as observed in Refs. \cite{gong14,schachenmayer13,pupillo14}).  We also note that, while we have ruled out the possibility of a logarithmic light cone in favor of one that is a nearly-linear polynomial, it is possible that \emph{any} sub-linearity of the light cone is impossible above some critical $\alpha$.

We thank S.\ Michalakis, C.\ Monroe, L.-M.\ Duan, C.\ Senko, P.\ Richerme, M.\ Maghrebi,
A.\ J.\ Daley, M.\ Kastner, J.\,Schachenmayer, A.\,Lee, J.\,Smith, S.\,Manmana  K.\ R.\ A.\ Hazzard, A.\ M.\ Rey, G.\ Pupillo, D.\ Vodola, L.\ Lepori, and E.\ Ercolessi for
discussions.  This work was supported by the NSF PFC at the JQI, NSF PIF, the ARL, and the ARO. MFF thanks the NRC for support.

%%%%%%%%%% BEGIN: BIBLIOGRAPHY %%%%%%%%%%


%%%%%%%%%% END: BIBLIOGRAPHY %%%%%%%%%%

\begin{widetext}

\appendix

\renewcommand{\thesection}{S\arabic{section}} 
\renewcommand{\theequation}{S\arabic{equation}}
\renewcommand{\thefigure}{S\arabic{figure}}
\setcounter{equation}{0}
\setcounter{figure}{0}

\section*{Supplemental Material}

\subsection{Finite-range Lieb-Robinson bound for interaction-picture operators}

Interaction-picture operators, which evolve under a finite-range Hamiltonian, can be constrained by a standard Lieb-Robinson bound.  Here we present a useful form of this bound, and then show how it can be used to write an interaction picture operator, such as $\mc{A}(t)$ (which in general is not supported on any finite subset of the lattice at $t>0$) as the sum of a sequence of operators with linearly increasing (finite) support but exponentially decreasing norm.

\

Interaction-picture operators evolve under the influence of
\begin{equation}
H^{\rm sr}=\frac{1}{2}\sum_{\mu,y,z}J_{\mu}^{\rm sr}(y,z)V_{y\mu}V_{z\mu},
\end{equation}
which is a finite-range Hamiltonian with range $\chi$.  Our first goal is to derive a Lieb-Robinson bound for the commutator $[\mc{A}(t),O]$, where $O$ must be allowed (for the sake of later applications) to have an arbitrary support comprised of all lattice sites in some set $\mathscr{S}$.  Note that $\mc{A}(0)=A$ is still assumed to be supported on a single site $i$, and we will let $r$ be the minimum distance between $i$ and any site in $\mathscr{S}$.  Defining $J^{\rm sr}(y,z)\equiv \sum_{\mu}J^{\rm sr}_{\mu}(y,z)$ and $\lambda^{\rm sr}=\sum_{z}J^{\rm sr}(y,z)$, then replacing $J^{\rm sr}(y,y)\rightarrow\lambda^{\rm sr}$ (instead of $J^{\rm sr}(y,y)=0$), and defining
\begin{equation}
\mc{J}^{\rm sr}_a(i,j)=\sum_{z_1,\dots,z_{a-1}}J^{\rm sr}(i,z_1)J^{\rm sr}(z_1,z_2)\dots J^{\rm sr}(z_{a-1},j)
\end{equation}
[with the understanding that $\mathcal{J}^{\rm sr}_1(i,j)=J^{\rm sr}(i,j)$], a standard Lieb-Robinson bound for this Hamiltonian yields
\begin{align}
\lVert [\mc{A}(t),O]\rVert&\leq2\lVert A\rVert \lVert O\rVert\sum_{j\in\mathscr{S}}\sum_{a=\lceil r/\chi\rceil }^{\infty}\frac{(2 t)^a}{a!}\mc{J}^{\rm sr}_a(i,j)\nonumber\\
&\leq2\lVert A\rVert \lVert O\rVert\sum_{a=\lceil r/\chi \rceil }^{\infty}\frac{(4\lambda^{\rm sr} t)^a}{a!}\nonumber\\
\label{supeq:sr_lr_bound}
&\leq 2\lVert A\rVert \lVert O\rVert e^{vt-r/\chi },
\end{align}
where $v=4e\lambda^{\rm sr}$.  Thus time evolution under the short-range Hamiltonian gives rise to a light cone at $r = R\equiv \chi \times vt$, and an exponential decay (on the length scale $\chi $) outside of that light cone.

\

Now we would like to express $\mc{A}(t)$ as the sum of a sequence of operators, each having a finite support larger than the last, but a smaller norm.  To accomplish this, we first define $\mathscr{B}_{ l}(i)$ ($ l=0,1,2,3,\dots$) to be a ball centered on site $i$ of radius $R+ l\chi $, the complements $\overline{\mathscr{B}}_{ l}(i)$, and define
\begin{equation}
\mc{A}( l,t)=\int_{\overline{\mathscr{B}}_{ l}(i)}d\mu(U)U\mc{A}(t)U^{\dagger},
\end{equation}
where the integration is with respect to the Haar measure \cite{bravyi06}.  The operator $\mc{A}( l,t)$ is supported entirely (without approximation) on $\mathscr{B}_{ l}(i)$.  Now we can use \eref{supeq:sr_lr_bound} to write
\begin{align}
\lVert\mc{A}( l,t)-\mc{A}(t)\rVert&=\Big\lVert\int_{\overline{\mathscr{B}}_{ l}(i)}\!\!\!\!\!\!\!\!d\mu(U)(U\mc{A}(t)U^{\dagger}-\mc{A}(t)UU^{\dagger})\Big\rVert\nonumber\\
&\leq\int_{\overline{\mathscr{B}}_{ l}(i)}d\mu(U)\lVert[U,\mc{A}(t)]\rVert\leq 2\lVert A\rVert e^{-l}.
\end{align}
Now consider the identity
\begin{equation}
\mc{A}(m,t)=\mc{A}(0,t)+\sum_{ l=1}^{m}(\mc{A}( l,t)-\mc{A}( l-1,t)).
\end{equation}
From above we know that $\lim_{m\rightarrow\infty}\lVert\mc{A}(m,t)-\mc{A}(t)\rVert=0$, and so $\lim_{m\rightarrow\infty}\mc{A}(m,t)=\mc{A}(t)$.  Defining $\mc{A}^{0}(t)=\mc{A}(0,t)$ and $\mc{A}^{l>0}(t)\equiv\mc{A}(l,t)-\mc{A}(l-1,t)$, we have
\begin{equation}
\mc{A}(t)=\sum_{ l=0}^{\infty}\mc{A}^{ l}(t).
\end{equation}
Note that $\mc{A}^{ l}(t)$ is supported on $\mathscr{B}_{ l}(i)$, and we can use a triangle inequality to write
\begin{align}
\lVert\mc{A}^{ l}(t)\rVert&\leq 2\lVert A\rVert(1+e)e^{- l}\equiv c\lVert A\rVert e^{- l}.
\end{align}
Thus we have found a way to write $\mc{A}(t)$ as a sum of a sequence of operators $\mc{A}^{ l}(t)$ with increasing (but always finite) support and exponentially decreasing norm. 

\subsection{Treating dynamics due to the (interaction-picture) time-evolution operator $\mc{U}(t)$.}

The commutator bounded in \eref{supeq:sr_lr_bound} applies to the interaction-picture operator $\mc{A}(t)$, but ultimately we would like to derive a bound for the true (Heisenberg picture) evolution, which in the interaction picture can be written
\begin{align}
C_r(t)=\lVert[A(t),B]\rVert&=\lVert[\mc{A}(t),\mc{U}(t)B\mc{U}^{\dagger}(t)]\rVert\leq\sum_{ l=0}^{\infty}\lVert[\mc{A}^ l(t),\mc{U}(t)B\mc{U}^{\dagger}(t)]\rVert\equiv\sum_{ l=0}^{\infty}C^{l}_r(t).
\end{align}
In this section, we generalize the techniques originally employed by Lieb and Robinson in order to bound commutator norms of the form $C^{l}_r(t)$.  The overall strategy is to use the interaction-picture equation of motion for $\mc{U}(t)$ to write a first-order differential equation for the relevant commutator, and then to convert this into an integral equation that can be iterated to produce the series presented in \eref{eq:prelim_bound} of the manuscript.

\

To begin, we introduce the generalized (two-time) commutator,
\begin{align}
\label{sup:starting_point}
\mc{C}^l_r(t,\tau)=[\mc{A}^ l(t),\mc{U}(\tau)B\mc{U}^{\dagger}(\tau)],
\end{align}
such that $C^l_r(t)=\lVert \mc{C}^l_r(t,t)\rVert$.  Using the equation of motion $d\,\mc{U}(\tau)/d\tau=-i\mc{H}^{\rm lr}(\tau)\mc{U}(\tau)$ and a Jacobi identity we can write
\begin{align}
\label{supeq:first_dif_eq}
\frac{d}{d\tau}\mc{C}^l_r(t,\tau)=-i\left([\mc{H}^{\rm lr}(\tau),\mc{C}^l_r(t,\tau)]-[\mc{U}(\tau)B\mc{U}^{\dagger}(\tau),[\mc{A}^{ l}(t),\mc{H}^{\rm lr}(\tau)]]\right).
\end{align}
Focusing on the interior commutator of the second term, defining the collective index $\xi=\{y,z,m,n\}$, and
\begin{align}
\mc{W}_{\xi}(\tau)\equiv \frac{1}{2}\sum_{\mu}J_{\mu}^{\rm lr}(y,z)\mc{V}^{m}_{y\mu}(\tau)\mc{V}^{n}_{z\mu}(\tau),
\end{align}
we have
\begin{align}
\label{supeq:H_reduce}
[\mc{A}^{ l}(t),\mc{H}^{\rm lr}(\tau)]&=\sum_{\xi}[\mc{A}^{ l}(t),\mc{W}_{\xi}(\tau)]=\sum_{\xi}[\mc{A}^{ l}(t),\mc{W}_{\xi}(\tau)\tilde{D}_{\rm i}(\xi)]=[\mc{A}^{ l}(t),\tilde{\mc{H}}^{\rm lr}(\tau)],
\end{align}
where
\begin{equation}
\tilde{\mc{H}}^{\rm lr}(\tau)=\sum_{\xi}\mc{W}_{\xi}(\tau)\tilde{D}_{\rm i}(\xi).
\end{equation}
Here the function $\tilde{D}_{\rm i}(\xi)$ is defined such that
\begin{equation}
\tilde{D}_{\rm i}(\xi)=\left\{\begin{array}{cl}
      1 & ~{\rm if}~~\mathscr{B}_{ l}(i)\cap[\mathscr{B}_{m}(y)\cup \mathscr{B}_n(z)]\neq\varnothing\\
      0 & ~{\rm otherwise},\\
\end{array}\right.
\end{equation}
i.e.~it is unity whenever the support of $\mc{A}^{l}(t)$ overlaps the support of $\mc{W}_{\xi}(t)$ (which encompasses the support of $\mc{W}_{\xi}(\tau\leq t)$).  Substituting \eref{supeq:H_reduce} into \eref{supeq:first_dif_eq} and again using a Jacobi identity, we obtain
\begin{align}
\label{supeq:apply_lemma}
\frac{d}{d\tau}\mc{C}^l_r(t,\tau)=-i[\mc{H}^{\rm lr}(\tau)-\tilde{\mc{H}}^{\rm lr}(\tau),\mc{C}^l_r(t,\tau)]-i[\mc{A}^{ l}(t),[\tilde{\mc{H}}^{\rm lr}(\tau),\mc{U}(\tau)B\mc{U}^{\dagger}(\tau)].
\end{align}
We now employ a result valid for differential equations of the form $d\mc{O}(\tau)/d\tau=i[\mc{H}(\tau),\mc{O}(\tau)]+\mc{G}(\tau)$ with $\mc{H}(\tau)$ any Hermitian operator, 
\begin{equation}
\label{supeq:technical_ODE_result}
\lVert\mc{O}(t)\rVert\leq\lVert\mc{O}(0)\rVert+\int_{0}^{t}d\tau\lVert \mc{G}(\tau)\rVert.
\end{equation}
Applying \eref{supeq:technical_ODE_result} to the differential equation in \eref{supeq:apply_lemma}, we obtain
\begin{align}
\label{eq:first_order}
C_{r}^l(t)&=\lVert \mc{C}_{r}^{l}(t,t)\rVert\leq\lVert[\mc{A}^{ l}(t),B]\rVert+2\lVert \mc{A}^{ l}(t)\rVert\sum_{\xi}\tilde{D}_{\rm i}(\xi)\int_{0}^{t} d\tau\lVert[\mc{W}_{\xi}(\tau),\mc{U}(\tau)B\mc{U}^{\dagger}(\tau)]\rVert.
\end{align}
The commutator norm inside the integral is precisely of the form $C^l_{r}(t)$, except that $t$ has been replaced with $\tau$ and $\mc{A}^{ l}(\tau)$ has been replaced with $\mc{W}_{\xi}(\tau)$.  Therefore the procedure just followed can be repeated, yielding
\begin{align}
\lVert[\mc{W}_{\xi}(\tau),\mc{U}(\tau)B\mc{U}^{\dagger}(\tau)]\rVert&\leq\lVert[\mc{W}_{\xi}(\tau),B]\rVert+2\lVert\mc{W}_{\xi}(\tau)\rVert\sum_{\xi'}\tilde{D}(\xi,\xi')\int_{0}^{\tau} d\tau'\lVert[\mc{W}_{\xi'}(\tau'),\mc{U}(\tau')B\mc{U}^{\dagger}(\tau')]\rVert,
\end{align}
where now
\begin{equation}
\tilde{D}(\xi,\xi')=\left\{\begin{array}{cl}
      1 & ~{\rm if}~~[\mathscr{B}_{m}(y)\cup\mathscr{B}_{n}(z)]\cap[\mathscr{B}_{m'}(y')\cup \mathscr{B}_{n'}(z')]\neq\varnothing\\
      0 & ~{\rm otherwise}.\\
\end{array}\right.
\end{equation}
Plugging this result back into Eq.\,\eqref{eq:first_order}, and then iterating this procedure repeatedly while taking advantage of the fact that $\lVert \mc{W}(\tau)\rVert$ does not actually depend on time, we obtain
\begin{align}
C_{r}^l(t)&\leq\lVert[\mc{A}^{l}(t),B]\rVert+2\lVert \mc{A}^{ l}(t)\rVert\lVert B\rVert\sum_{\xi_1}\tilde{D}_{\rm i}(\xi_1)\lVert \mc{W}_{\xi_1}\rVert \tilde{D}_{\rm f}(\xi_1)\times  2\int_{0}^{t} d\tau_1\\
&+2\lVert \mc{A}^{ l}(t)\rVert\lVert B\rVert\times\sum_{\xi_1,\xi_2}\tilde{D}_{\rm i}(\xi_1)\lVert\mc{W}_{\xi_1}\rVert \tilde{D}(\xi_1,\xi_2)\lVert\mc{W}_{\xi_2}\rVert \tilde{D}_{\rm f}(\xi_2)\times2^2\int_{0}^{t} d\tau_1\int_{0}^{\tau_1} d\tau_2\\
&+2\lVert \mc{A}^{ l}(t)\rVert\lVert B\rVert\times\sum_{\xi_1,\xi_2,\xi_3}\tilde{D}_{\rm i}(\xi_1)\lVert\mc{W}_{\xi_1}\rVert \tilde{D}(\xi_1,\xi_2)\lVert\mc{W}_{\xi_2}\rVert \tilde{D}(\xi_2,\xi_3)\lVert\mc{W}_{\xi_3}\rVert \tilde{D}_{\rm f}(\xi_3)\times2^3\int_{0}^{t} d\tau_1\int_{0}^{\tau_1} d\tau_2\int_{0}^{\tau_2}d\tau_3\\
&+\dots\nonumber
\end{align}
From the definition, the function $\tilde{D}(\xi_1,\xi_2)$ vanishes unless the sets $\mathscr{B}_{m_1}(y_1)\cup\mathscr{B}_{n_1}(z_1)$ and $\mathscr{B}_{m_2}(y_2)\cup\mathscr{B}_{n_2}(z_2)$ overlap.  It is straightforward (though requires some careful inspection) to check that the replacement of $\tilde{D}(\xi_1,\xi_2)$ by $2D(\xi_1,\xi_2)$, where
\begin{align}
D(\xi_1,\xi_2)=\left\{\begin{array}{cl}
      1 &~{\rm if}~~ \mathscr{B}_{n_1}(z_1)\cap\mathscr{B}_{m_2}(y_2)\neq \varnothing\\
      0 & ~{\rm otherwise},\\
\end{array}\right.
\end{align}
renders the summation strictly larger, so long as we replace $J^{\rm lr}(y,y)\rightarrow \kappa\lambda_{\chi}$ (remember it is zero by our previous definition), where $\kappa=\sum_{l=0}^{\infty}e^{-l}=e/(e-1)$ and $\lambda_{\chi}=\sum_{z}J^{\rm lr}(y,z)=\lambda\times\chi^{D-\alpha}$ (with $\lambda$ an order unity constant).  This is possible because whenever $\mathscr{B}_{m_2}(y_2)$ overlaps with $\mathscr{B}_{m_1}(y_1)$ instead of with $\mathscr{B}_{n_1}(z_1)$, such that the endpoint $z_1$ and its support $n_1$ can be freely summed over, we can pretend that $J^{\rm lr}(y_1,z_1)$ always returns to where it started from [i.e. it gets replaced with $J^{\rm lr}(y_1,y_1)$] but with the value of the free summation over $z_1$ and $n_1$ ($\kappa\lambda_{\chi}$).  Carrying out the time integrals and summing over $l$ then gives us \eref{eq:prelim_bound} and \eref{eq:J_def} of the main text.

\subsection{Reduction of $\mc{J}_{a}(i,j)$ to a power law}

Equations (\ref{eq:prelim_bound}) and (\ref{eq:J_def}) of the manuscript already comprise an effective bound on the dynamics (note that these expressions are just sums over expectation values that all have trivial bounds, and there are no operators left).  However, the series is rather complicated, and without further simplification it is difficult to extract a light-cone shape.  Here we show how to reduce this series to a much simpler (though strictly looser) bound, with a functional form that makes the light-cone shape easy to extract.

\

To begin, we use the inequality $\lVert\mc{W}_{\xi}\rVert\leq c^2 J^{\rm lr}(y,z)e^{-(m+n)}/2$ to bound the expressions in \eref{eq:J_def} as
\begin{align}
\mc{J}_{a}(i,j)&\leq (2c^2)^{a}\sum_{l}\sum_{\xi_1,\dots,\xi_a}e^{- l}D_{\rm i}(\xi_1)e^{-(m_1+n_1)}J^{\rm lr}(y_1,z_1)D(\xi_1,\xi_2)e^{-(m_2+n_2)}J^{\rm lr}(y_2,z_2)\times\dots\nonumber\\
&\dots\times D(\xi_{a-1},\xi_a)e^{-(m_a+n_a)}J^{\rm lr}(y_a,z_a)D_{\rm f}(\xi_a).
\label{sup:D_contract}
\end{align}
Now we can carry out the summations over indices $m$ and $n$ one by one.  In particular, it is straightforward to show that $\sum_{n_1,m_2}e^{-n_1}D(\xi_1,\xi_2)e^{-m_2}\leq \kappa^2 K(z_1,y_2)$, with
\begin{align}
K(z,y)=\left\{\begin{array}{cc}
      1 &~{\rm if}~~d(z,y)\leq2R\\
  e^{-[(d(z,y)-2R)/2\chi ]} & ~{\rm if}~~d(z,y)>2R.\\
\end{array}\right.
\end{align}
Using this result repeatedly in \eref{sup:D_contract} gives \eref{eq:JK_def} of the main text.

\

The next step is to carry out the summations over indices $y$, which requires bounding the discrete convolution
\begin{align}
\label{supeq:KJF}
\sum_{y_2}K(z_1,y_2)J^{\rm lr}(y_2,z_2)\leq (2\kappa\lambda_{\chi})\times F(z_1,z_2)
\end{align}
(the prefactor $2\kappa\lambda_{\chi}$ is a matter of definition, but is convenient in what follows).  It's worth noting that, because $K(z_1,y_2)$ decays exponentially in $d(z_1,y_2)$ while $J^{\rm lr}(y_2,z_2)$ decays algebraically in $d(y_2,z_2)$, we can already anticipate that the convolution will be dominated (at large distances) by terms where the intermediate point $y_2$ is much closer to $z_1$ than to $z_2$, and thus $F(z_1,z_2)$ should decay algebraically at large $d(z_1,z_2)$.  To explicitly construct a valid function $F$, we first note that a trivial bound can be obtained at any value of $r$ by simply using $K(z_1,y_2)\leq 1$, which leads to
\begin{equation}
\label{eq:fdj}
\sum_{y_2}K(z_1,y_2)J^{\rm lr}(y_2,z_2)\leq (\kappa+1)\lambda_{\chi}\leq 2\kappa\lambda_{\chi}.
\end{equation}
When $R\gg \chi $ (equivalent to $vt\gg 1$), which is the situation we will eventually consider, we actually expect this bound to be tight for $r\lesssim R$.  Recovering an algebraic decay when $r\gtrsim R$, which is the regime we will eventually be interested in, is more challenging, and can be carried out as follows (Fig. \ref{supfig:contraction} summarizes the following discussion).  First, we imagine a line connecting points $z_1$ and $z_2$, with length $d(z_1,z_2)$, and let the point $w$ sit along that line at a distance $[d(z_1,z_2)-2R]/2$ from $z_2$ (see the figure below).  For $d(z_1,z_2)>6R$, it is straightforward to show that $d(z_1,w)-2R>d(z_1,z_2)/3$ and that $d(w,z_2)>d(z_1,z_2)/3$.  We now imagine a plane passing through the point $w$ and perpendicular to the line separating $z_1$ from $z_2$, and denote the half-space on the $z_{1(2)}$ side of that plane by $\mc{Z}_{1(2)}$.  In the sum in Eq.\,\eqref{eq:fdj}, any intermediate point $y_2$ is either in $\mc{Z}_1$ or $\mc{Z}_2$ (or in both if it lies right in the plane separating $\mc{Z}_1$ from $\mc{Z}_2$).  We can therefore write
\begin{align}
\sum_{y_2}K(z_1,y_2)J^{\rm lr}(y_2,z_2)&\leq \sum_{y_2\in \mc{Z}_1}  K(z_1,y_2)J^{\rm lr}(y_2,z_2)+ \sum_{y_2 \in \mc{Z}_2 } K(z_1,y_2)J^{\rm lr}(y_2,z_2)\\
&\leq \frac{1}{(d(z_1,z_2)/3)^{\alpha}}\sum_{y_2}K(z_1,y_2)+e^{-d(z_1,z_2)/(6\chi )}\sum_{y_2}J^{\rm lr}(y_2,z_2)\\
&=\frac{b R^{D}}{(2\chi )^{\alpha}}\frac{1}{\delta^{\alpha}}+2\kappa\lambda_{\chi} e^{-\delta},
\end{align}
where $b$ is an order unity constant and $\delta=d(z_1,z_2)/(6\chi )$.  Now it is straightforward to show that the final line satisfies
\begin{equation}
\frac{b R^{D}}{(2\chi )^{\alpha}}\frac{1}{\delta^{\alpha}}+2\kappa\lambda_{\chi} e^{-\delta}\leq 2\frac{b R^{D}}{(2\chi )^{\alpha}}\frac{1}{\delta^{\alpha}}= \frac{2b  R^{D}3^{\alpha}}{d(z_1,z_2)^{\alpha}}
\label{sup:almost_contracted}
\end{equation}
whenever $\delta>\alpha\log\alpha$, or alternatively when $d(z_1,z_2)\geq 6\chi \alpha\log\alpha$.  Strictly speaking this is only guaranteed when $t$ is sufficiently large compared to unity (by an amount that depends on $\alpha$), in which case $b R^D/(4\chi )^{\alpha}\sim t^D\lambda^{\rm lr}$ will be larger than $2\lambda^{\rm lr}$.  However, we are ultimately interested in the large $t$ limit, so this is fine.  Since $R\propto \chi  t$, and we are interested in the large $t$ limit, for any finite $\alpha$ this condition is already guaranteed by the condition $d(z_1,z_2)>6R$, but strictly speaking we should remember that our final results are valid whenever $6R>6\chi\alpha\log\alpha$, or $vt>\alpha\log\alpha$.  Thus we arrive at a valid bound using
\begin{align}
F(z_1,z_2)=\left\{\begin{array}{cc}
      1 &~{\rm if}~~d(z_1,z_2)\leq6R\\
(6R/d(z_1,z_2))^{\alpha} & ~{\rm if}~~d(z_1,z_2)>6R.\\
\end{array}\right.
\end{align}
Note that the second line is obtained by first multiplying the right-hand side of the bound in \eref{sup:almost_contracted} by $2^{\alpha}R^{\alpha-D}\kappa\lambda_{\chi}/b\gg1$ in order to make $F$ continuous at $d(z_1,z_2)=6R$.
\begin{figure}[t!]
\centering
\includegraphics[width=10cm]{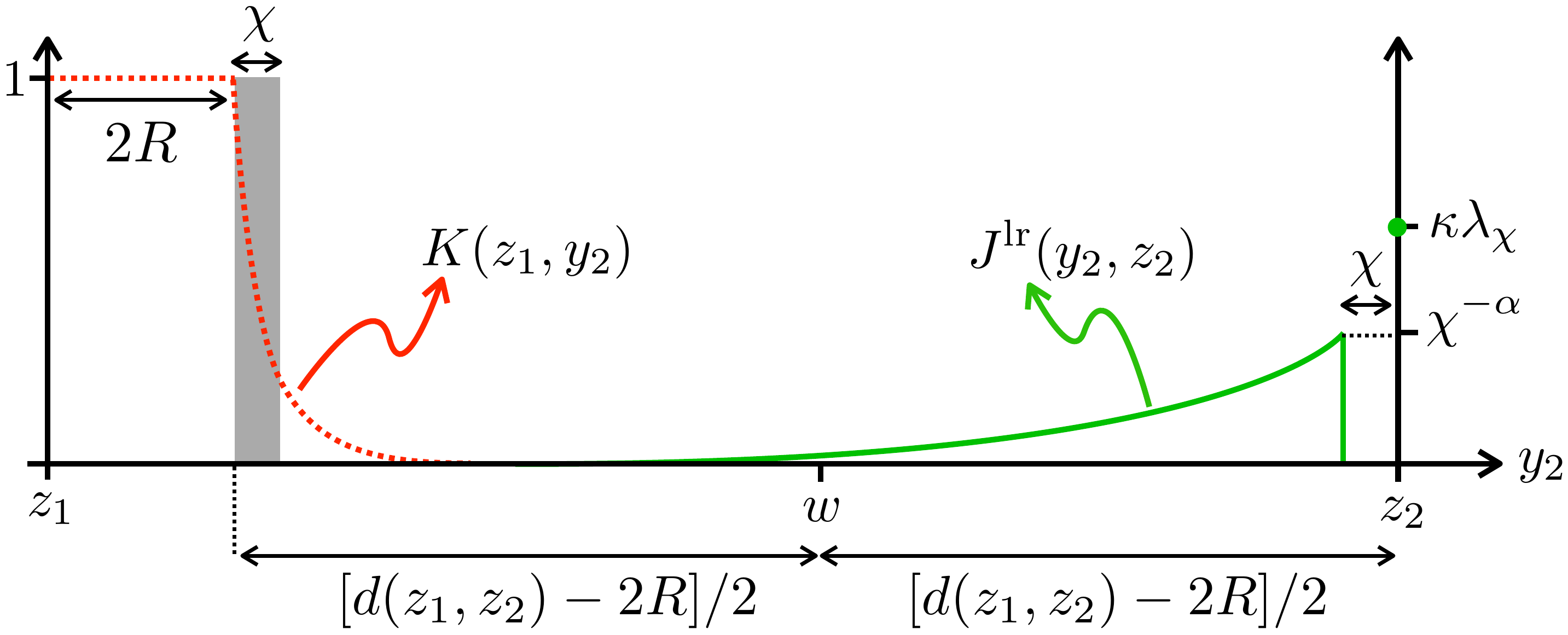}
\caption{Schematic summarizing the discrete convolution of $K(z_1,y_2)$ with $J^{\rm lr}(y_2,z_2)$. }
\label{supfig:contraction}
\end{figure}
Plugging \eref{supeq:KJF} into \eref{eq:just_Fs} of the main text, and utilizing the inequality $K(z_a,j)\leq 2F(z_a,j)$ [which can be derived by analyzing just the term where $y_2=z_2$ on the left-hand side of \eref{supeq:KJF}], we can now write
\begin{align}
\mc{J}_{a}(i,j)\leq 2\kappa^2(4\lambda_{\chi}\kappa^3 c^2)^a  \sum_{z_1,\dots,z_a}F(i,z_1)F(z_1,z_2)\times\dots\times F(z_{a-1},z_a)F(z_a,j).
\end{align}
The $F$'s now obey a simple reproducibility condition \cite{Hastings05},
\begin{equation}
\sum_{z_2}F(z_1,z_2)F(z_2,z_3)\leq g R^{D}F(z_1,z_3),
\end{equation}
where $g$ is a constant.  Collecting our many constants into $\vartheta=g\times 4\kappa^3 c^2$, defining the cutoff-dependent (and time-dependent) velocity $v_{\chi}=\vartheta\times(R^{D}\lambda_{\chi})$, and absorbing the constant factor $2\times 6^{\alpha}$ into $\kappa^2$, we then have (for $r>6R$)
\begin{align}
\mc{J}_{a}(i,j)&\leq \kappa^2\left(R/r\right)^{\alpha}\times (v_{\chi})^a.
\end{align}
Plugging this back into \eref{eq:prelim_bound} of the main text, we obtain our final bound [\eref{eq:final_bound} of the main text],
\begin{align}
C_{r}(t)&\leq \sum_{\ell}\lVert[\mc{A}^{\ell}(t),B]\rVert+4c\kappa^2 (R/r)^{\alpha} \sum_{a=1}^{\infty}\frac{(v_{\chi}t)^a}{a!}\\
&\leq 2c\kappa\!\left(e^{vt-r/\chi}+2\kappa \frac{\exp[v_{\chi}t]-1}{(r/R)^{\alpha}}\right).
\end{align}

%
%
%\bibliographystyle{apsrev}
%\bibliography{refs}

\begin{thebibliography}{26}%
\makeatletter
\providecommand \@ifxundefined [1]{%
 \@ifx{#1\undefined}
}%
\providecommand \@ifnum [1]{%
 \ifnum #1\expandafter \@firstoftwo
 \else \expandafter \@secondoftwo
 \fi
}%
\providecommand \@ifx [1]{%
 \ifx #1\expandafter \@firstoftwo
 \else \expandafter \@secondoftwo
 \fi
}%
\providecommand \natexlab [1]{#1}%
\providecommand \enquote  [1]{``#1''}%
\providecommand \bibnamefont  [1]{#1}%
\providecommand \bibfnamefont [1]{#1}%
\providecommand \citenamefont [1]{#1}%
\providecommand \href@noop [0]{\@secondoftwo}%
\providecommand \href [0]{\begingroup \@sanitize@url \@href}%
\providecommand \@href[1]{\@@startlink{#1}\@@href}%
\providecommand \@@href[1]{\endgroup#1\@@endlink}%
\providecommand \@sanitize@url [0]{\catcode `\\12\catcode `\$12\catcode
  `\&12\catcode `\#12\catcode `\^12\catcode `\_12\catcode `\%12\relax}%
\providecommand \@@startlink[1]{}%
\providecommand \@@endlink[0]{}%
\providecommand \url  [0]{\begingroup\@sanitize@url \@url }%
\providecommand \@url [1]{\endgroup\@href {#1}{\urlprefix }}%
\providecommand \urlprefix  [0]{URL }%
\providecommand \Eprint [0]{\href }%
\providecommand \doibase [0]{http://dx.doi.org/}%
\providecommand \selectlanguage [0]{\@gobble}%
\providecommand \bibinfo  [0]{\@secondoftwo}%
\providecommand \bibfield  [0]{\@secondoftwo}%
\providecommand \translation [1]{[#1]}%
\providecommand \BibitemOpen [0]{}%
\providecommand \bibitemStop [0]{}%
\providecommand \bibitemNoStop [0]{.\EOS\space}%
\providecommand \EOS [0]{\spacefactor3000\relax}%
\providecommand \BibitemShut  [1]{\csname bibitem#1\endcsname}%
\let\auto@bib@innerbib\@empty
%</preamble>
\bibitem [{\citenamefont {Lieb}\ and\ \citenamefont {Robinson}(1972)}]{lieb72}%
  \BibitemOpen
  \bibfield  {author} {\bibinfo {author} {\bibfnamefont {E.~H.}\ \bibnamefont
  {Lieb}}\ and\ \bibinfo {author} {\bibfnamefont {D.~W.}\ \bibnamefont
  {Robinson}},\ }\href@noop {} {\bibfield  {journal} {\bibinfo  {journal}
  {Comm. Math. Phys.}\ }\textbf {\bibinfo {volume} {28}},\ \bibinfo {pages}
  {251} (\bibinfo {year} {1972})}\BibitemShut {NoStop}%
\bibitem [{\citenamefont {Binder}\ and\ \citenamefont
  {Young}(1986)}]{RevModPhys.58.801}%
  \BibitemOpen
  \bibfield  {author} {\bibinfo {author} {\bibfnamefont {K.}~\bibnamefont
  {Binder}}\ and\ \bibinfo {author} {\bibfnamefont {A.~P.}\ \bibnamefont
  {Young}},\ }\href {\doibase 10.1103/RevModPhys.58.801} {\bibfield  {journal}
  {\bibinfo  {journal} {Rev. Mod. Phys.}\ }\textbf {\bibinfo {volume} {58}},\
  \bibinfo {pages} {801} (\bibinfo {year} {1986})}\BibitemShut {NoStop}%
\bibitem [{\citenamefont {Ruderman}\ and\ \citenamefont
  {Kittel}(1954)}]{PhysRev.96.99}%
  \BibitemOpen
  \bibfield  {author} {\bibinfo {author} {\bibfnamefont {M.~A.}\ \bibnamefont
  {Ruderman}}\ and\ \bibinfo {author} {\bibfnamefont {C.}~\bibnamefont
  {Kittel}},\ }\href {\doibase 10.1103/PhysRev.96.99} {\bibfield  {journal}
  {\bibinfo  {journal} {Phys. Rev.}\ }\textbf {\bibinfo {volume} {96}},\
  \bibinfo {pages} {99} (\bibinfo {year} {1954})}\BibitemShut {NoStop}%
\bibitem [{\citenamefont {Saffman}\ \emph {et~al.}(2010)\citenamefont
  {Saffman}, \citenamefont {Walker},\ and\ \citenamefont
  {M{\o}lmer}}]{saffman10}%
  \BibitemOpen
  \bibfield  {author} {\bibinfo {author} {\bibfnamefont {M.}~\bibnamefont
  {Saffman}}, \bibinfo {author} {\bibfnamefont {T.~G.}\ \bibnamefont {Walker}},
  \ and\ \bibinfo {author} {\bibfnamefont {K.}~\bibnamefont {M{\o}lmer}},\
  }\href@noop {} {\bibfield  {journal} {\bibinfo  {journal} {Rev. Mod. Phys.}\
  }\textbf {\bibinfo {volume} {82}},\ \bibinfo {pages} {2313} (\bibinfo {year}
  {2010})}\BibitemShut {NoStop}%
\bibitem [{\citenamefont {Islam}\ \emph {et~al.}(2013)\citenamefont {Islam},
  \citenamefont {Senko}, \citenamefont {Campbell}, \citenamefont {Korenblit},
  \citenamefont {Smith}, \citenamefont {Lee}, \citenamefont {Edwards},
  \citenamefont {Wang}, \citenamefont {Freericks},\ and\ \citenamefont
  {Monroe}}]{islam13}%
  \BibitemOpen
  \bibfield  {author} {\bibinfo {author} {\bibfnamefont {R.}~\bibnamefont
  {Islam}}, \bibinfo {author} {\bibfnamefont {C.}~\bibnamefont {Senko}},
  \bibinfo {author} {\bibfnamefont {W.~C.}\ \bibnamefont {Campbell}}, \bibinfo
  {author} {\bibfnamefont {S.}~\bibnamefont {Korenblit}}, \bibinfo {author}
  {\bibfnamefont {J.}~\bibnamefont {Smith}}, \bibinfo {author} {\bibfnamefont
  {A.}~\bibnamefont {Lee}}, \bibinfo {author} {\bibfnamefont {E.~E.}\
  \bibnamefont {Edwards}}, \bibinfo {author} {\bibfnamefont {C.-C.~J.}\
  \bibnamefont {Wang}}, \bibinfo {author} {\bibfnamefont {J.~K.}\ \bibnamefont
  {Freericks}}, \ and\ \bibinfo {author} {\bibfnamefont {C.}~\bibnamefont
  {Monroe}},\ }\href@noop {} {\bibfield  {journal} {\bibinfo  {journal}
  {Science}\ }\textbf {\bibinfo {volume} {340}},\ \bibinfo {pages} {583}
  (\bibinfo {year} {2013})}\BibitemShut {NoStop}%
\bibitem [{\citenamefont {Gopalakrishnan}\ \emph {et~al.}(2011)\citenamefont
  {Gopalakrishnan}, \citenamefont {Lev},\ and\ \citenamefont
  {Goldbart}}]{gopalakrishnan11}%
  \BibitemOpen
  \bibfield  {author} {\bibinfo {author} {\bibfnamefont {S.}~\bibnamefont
  {Gopalakrishnan}}, \bibinfo {author} {\bibfnamefont {B.~L.}\ \bibnamefont
  {Lev}}, \ and\ \bibinfo {author} {\bibfnamefont {P.~M.}\ \bibnamefont
  {Goldbart}},\ }\href@noop {} {\bibfield  {journal} {\bibinfo  {journal}
  {Phys. Rev. Lett.}\ }\textbf {\bibinfo {volume} {107}},\ \bibinfo {pages}
  {277201} (\bibinfo {year} {2011})}\BibitemShut {NoStop}%
\bibitem [{\citenamefont {Aikawa}\ \emph {et~al.}(2012)\citenamefont {Aikawa},
  \citenamefont {Frisch}, \citenamefont {Mark}, \citenamefont {Baier},
  \citenamefont {Rietzler}, \citenamefont {Grimm},\ and\ \citenamefont
  {Ferlaino}}]{aikawa12}%
  \BibitemOpen
  \bibfield  {author} {\bibinfo {author} {\bibfnamefont {K.}~\bibnamefont
  {Aikawa}}, \bibinfo {author} {\bibfnamefont {A.}~\bibnamefont {Frisch}},
  \bibinfo {author} {\bibfnamefont {M.}~\bibnamefont {Mark}}, \bibinfo {author}
  {\bibfnamefont {S.}~\bibnamefont {Baier}}, \bibinfo {author} {\bibfnamefont
  {A.}~\bibnamefont {Rietzler}}, \bibinfo {author} {\bibfnamefont
  {R.}~\bibnamefont {Grimm}}, \ and\ \bibinfo {author} {\bibfnamefont
  {F.}~\bibnamefont {Ferlaino}},\ }\href@noop {} {\bibfield  {journal}
  {\bibinfo  {journal} {Phys. Rev. Lett.}\ }\textbf {\bibinfo {volume} {108}},\
  \bibinfo {pages} {210401} (\bibinfo {year} {2012})}\BibitemShut {NoStop}%
\bibitem [{\citenamefont {Yan}\ \emph {et~al.}(2013)\citenamefont {Yan},
  \citenamefont {Moses}, \citenamefont {Gadway}, \citenamefont {Covey},
  \citenamefont {Hazzard}, \citenamefont {Rey}, \citenamefont {Jin},\ and\
  \citenamefont {Ye}}]{yan13}%
  \BibitemOpen
  \bibfield  {author} {\bibinfo {author} {\bibfnamefont {B.}~\bibnamefont
  {Yan}}, \bibinfo {author} {\bibfnamefont {S.~A.}\ \bibnamefont {Moses}},
  \bibinfo {author} {\bibfnamefont {B.}~\bibnamefont {Gadway}}, \bibinfo
  {author} {\bibfnamefont {J.~P.}\ \bibnamefont {Covey}}, \bibinfo {author}
  {\bibfnamefont {K.~R.~A.}\ \bibnamefont {Hazzard}}, \bibinfo {author}
  {\bibfnamefont {A.~M.}\ \bibnamefont {Rey}}, \bibinfo {author} {\bibfnamefont
  {D.~S.}\ \bibnamefont {Jin}}, \ and\ \bibinfo {author} {\bibfnamefont
  {J.}~\bibnamefont {Ye}},\ }\href@noop {} {\bibfield  {journal} {\bibinfo
  {journal} {Nature (London)}\ }\textbf {\bibinfo {volume} {501}},\ \bibinfo
  {pages} {521} (\bibinfo {year} {2013})}\BibitemShut {NoStop}%
\bibitem [{\citenamefont {Eisert}\ \emph {et~al.}(2013)\citenamefont {Eisert},
  \citenamefont {van~den Worm}, \citenamefont {Manmana},\ and\ \citenamefont
  {Kastner}}]{eisert13}%
  \BibitemOpen
  \bibfield  {author} {\bibinfo {author} {\bibfnamefont {J.}~\bibnamefont
  {Eisert}}, \bibinfo {author} {\bibfnamefont {M.}~\bibnamefont {van~den
  Worm}}, \bibinfo {author} {\bibfnamefont {S.~R.}\ \bibnamefont {Manmana}}, \
  and\ \bibinfo {author} {\bibfnamefont {M.}~\bibnamefont {Kastner}},\
  }\href@noop {} {\bibfield  {journal} {\bibinfo  {journal} {Phys. Rev. Lett.}\
  }\textbf {\bibinfo {volume} {111}},\ \bibinfo {pages} {260401} (\bibinfo
  {year} {2013})}\BibitemShut {NoStop}%
\bibitem [{\citenamefont {Hauke}\ and\ \citenamefont
  {Tagliacozzo}(2013)}]{hauke13}%
  \BibitemOpen
  \bibfield  {author} {\bibinfo {author} {\bibfnamefont {P.}~\bibnamefont
  {Hauke}}\ and\ \bibinfo {author} {\bibfnamefont {L.}~\bibnamefont
  {Tagliacozzo}},\ }\href@noop {} {\bibfield  {journal} {\bibinfo  {journal}
  {Phys. Rev. Lett.}\ }\textbf {\bibinfo {volume} {111}} (\bibinfo {year}
  {2013})}\BibitemShut {NoStop}%
\bibitem [{\citenamefont {Juenemann}\ \emph {et~al.}(2013)\citenamefont
  {Juenemann}, \citenamefont {Cadarso}, \citenamefont {Perez-Garcia},
  \citenamefont {Bermudez},\ and\ \citenamefont {Garcia-Ripoll}}]{juenemann13}%
  \BibitemOpen
  \bibfield  {author} {\bibinfo {author} {\bibfnamefont {J.}~\bibnamefont
  {Juenemann}}, \bibinfo {author} {\bibfnamefont {A.}~\bibnamefont {Cadarso}},
  \bibinfo {author} {\bibfnamefont {D.}~\bibnamefont {Perez-Garcia}}, \bibinfo
  {author} {\bibfnamefont {A.}~\bibnamefont {Bermudez}}, \ and\ \bibinfo
  {author} {\bibfnamefont {J.~J.}\ \bibnamefont {Garcia-Ripoll}},\ }\href@noop
  {} {\bibfield  {journal} {\bibinfo  {journal} {Phys. Rev. Lett.}\ }\textbf
  {\bibinfo {volume} {111}},\ \bibinfo {pages} {230404} (\bibinfo {year}
  {2013})}\BibitemShut {NoStop}%
\bibitem [{\citenamefont {Gong}\ \emph
  {et~al.}(2014{\natexlab{a}})\citenamefont {Gong}, \citenamefont {Foss-Feig},
  \citenamefont {Michalakis},\ and\ \citenamefont {Gorshkov}}]{gong14}%
  \BibitemOpen
  \bibfield  {author} {\bibinfo {author} {\bibfnamefont {Z.-X.}\ \bibnamefont
  {Gong}}, \bibinfo {author} {\bibfnamefont {M.}~\bibnamefont {Foss-Feig}},
  \bibinfo {author} {\bibfnamefont {S.}~\bibnamefont {Michalakis}}, \ and\
  \bibinfo {author} {\bibfnamefont {A.~V.}\ \bibnamefont {Gorshkov}},\ }\href
  {\doibase 10.1103/PhysRevLett.113.030602} {\bibfield  {journal} {\bibinfo
  {journal} {Phys. Rev. Lett.}\ }\textbf {\bibinfo {volume} {113}},\ \bibinfo
  {pages} {030602} (\bibinfo {year} {2014}{\natexlab{a}})}\BibitemShut
  {NoStop}%
\bibitem [{\citenamefont {Richerme}\ \emph {et~al.}(2014)\citenamefont
  {Richerme}, \citenamefont {Gong}, \citenamefont {Lee}, \citenamefont {Senko},
  \citenamefont {Smith}, \citenamefont {Foss-Feig}, \citenamefont {Michalakis},
  \citenamefont {Gorshkov},\ and\ \citenamefont {Monroe}}]{Richerme14}%
  \BibitemOpen
  \bibfield  {author} {\bibinfo {author} {\bibfnamefont {P.}~\bibnamefont
  {Richerme}}, \bibinfo {author} {\bibfnamefont {Z.-X.}\ \bibnamefont {Gong}},
  \bibinfo {author} {\bibfnamefont {A.}~\bibnamefont {Lee}}, \bibinfo {author}
  {\bibfnamefont {C.}~\bibnamefont {Senko}}, \bibinfo {author} {\bibfnamefont
  {J.}~\bibnamefont {Smith}}, \bibinfo {author} {\bibfnamefont
  {M.}~\bibnamefont {Foss-Feig}}, \bibinfo {author} {\bibfnamefont
  {S.}~\bibnamefont {Michalakis}}, \bibinfo {author} {\bibfnamefont {A.~V.}\
  \bibnamefont {Gorshkov}}, \ and\ \bibinfo {author} {\bibfnamefont
  {C.}~\bibnamefont {Monroe}},\ }\href@noop {} {\bibfield  {journal} {\bibinfo
  {journal} {arXiv:1401.5088}\ } (\bibinfo {year} {2014})}\BibitemShut
  {NoStop}%
\bibitem [{\citenamefont {Jurcevic}\ \emph {et~al.}(2014)\citenamefont
  {Jurcevic}, \citenamefont {Lanyon}, \citenamefont {Hauke}, \citenamefont
  {Hempel}, \citenamefont {Zoller}, \citenamefont {Blatt},\ and\ \citenamefont
  {Roos}}]{Jurcevic14}%
  \BibitemOpen
  \bibfield  {author} {\bibinfo {author} {\bibfnamefont {P.}~\bibnamefont
  {Jurcevic}}, \bibinfo {author} {\bibfnamefont {B.~P.}\ \bibnamefont
  {Lanyon}}, \bibinfo {author} {\bibfnamefont {P.}~\bibnamefont {Hauke}},
  \bibinfo {author} {\bibfnamefont {C.}~\bibnamefont {Hempel}}, \bibinfo
  {author} {\bibfnamefont {P.}~\bibnamefont {Zoller}}, \bibinfo {author}
  {\bibfnamefont {R.}~\bibnamefont {Blatt}}, \ and\ \bibinfo {author}
  {\bibfnamefont {C.~F.}\ \bibnamefont {Roos}},\ }\href@noop {} {\bibfield
  {journal} {\bibinfo  {journal} {arXiv:1401.5387}\ } (\bibinfo {year}
  {2014})}\BibitemShut {NoStop}%
\bibitem [{\citenamefont {Bose}(2007)}]{bose07}%
  \BibitemOpen
  \bibfield  {author} {\bibinfo {author} {\bibfnamefont {S.}~\bibnamefont
  {Bose}},\ }\href@noop {} {\bibfield  {journal} {\bibinfo  {journal} {Contemp.
  Phys.}\ }\textbf {\bibinfo {volume} {48}},\ \bibinfo {pages} {13} (\bibinfo
  {year} {2007})}\BibitemShut {NoStop}%
\bibitem [{\citenamefont {Polkovnikov}\ \emph {et~al.}(2011)\citenamefont
  {Polkovnikov}, \citenamefont {Sengupta}, \citenamefont {Silva},\ and\
  \citenamefont {Vengalattore}}]{polkovnikov11}%
  \BibitemOpen
  \bibfield  {author} {\bibinfo {author} {\bibfnamefont {A.}~\bibnamefont
  {Polkovnikov}}, \bibinfo {author} {\bibfnamefont {K.}~\bibnamefont
  {Sengupta}}, \bibinfo {author} {\bibfnamefont {A.}~\bibnamefont {Silva}}, \
  and\ \bibinfo {author} {\bibfnamefont {M.}~\bibnamefont {Vengalattore}},\
  }\href@noop {} {\bibfield  {journal} {\bibinfo  {journal} {Rev. Mod. Phys.}\
  }\textbf {\bibinfo {volume} {83}},\ \bibinfo {pages} {863} (\bibinfo {year}
  {2011})}\BibitemShut {NoStop}%
\bibitem [{\citenamefont {Hastings}\ and\ \citenamefont
  {Koma}(2006)}]{Hastings05}%
  \BibitemOpen
  \bibfield  {author} {\bibinfo {author} {\bibfnamefont {M.}~\bibnamefont
  {Hastings}}\ and\ \bibinfo {author} {\bibfnamefont {T.}~\bibnamefont
  {Koma}},\ }\href {\doibase 10.1007/s00220-006-0030-4} {\bibfield  {journal}
  {\bibinfo  {journal} {Comm. Math. Phys.}\ }\textbf {\bibinfo {volume}
  {265}},\ \bibinfo {pages} {781} (\bibinfo {year} {2006})}\BibitemShut
  {NoStop}%
\bibitem [{\citenamefont {Wolf}\ \emph {et~al.}(2008)\citenamefont {Wolf},
  \citenamefont {Verstraete}, \citenamefont {Hastings},\ and\ \citenamefont
  {Cirac}}]{PhysRevLett.100.070502}%
  \BibitemOpen
  \bibfield  {author} {\bibinfo {author} {\bibfnamefont {M.~M.}\ \bibnamefont
  {Wolf}}, \bibinfo {author} {\bibfnamefont {F.}~\bibnamefont {Verstraete}},
  \bibinfo {author} {\bibfnamefont {M.~B.}\ \bibnamefont {Hastings}}, \ and\
  \bibinfo {author} {\bibfnamefont {J.~I.}\ \bibnamefont {Cirac}},\ }\href
  {\doibase 10.1103/PhysRevLett.100.070502} {\bibfield  {journal} {\bibinfo
  {journal} {Phys. Rev. Lett.}\ }\textbf {\bibinfo {volume} {100}},\ \bibinfo
  {pages} {070502} (\bibinfo {year} {2008})}\BibitemShut {NoStop}%
\bibitem [{Note1()}]{Note1}%
  \BibitemOpen
  \bibinfo {note} {The results presented are significantly more general than
  Eq.\protect \tmspace +\thinmuskip {.1667em}\protect \textup {\hbox
  {\mathsurround \z@ \protect \normalfont (\ignorespaces \ref
  {eq:hamiltonian}\unskip \@@italiccorr )}} suggests, and can easily be
  generalized to fermionic models, or Hamiltonians with arbitrary
  single-particle terms or time dependence.}\BibitemShut {Stop}%
\bibitem [{\citenamefont {Bravyi}\ \emph {et~al.}(2006)\citenamefont {Bravyi},
  \citenamefont {Hastings},\ and\ \citenamefont {Verstraete}}]{bravyi06}%
  \BibitemOpen
  \bibfield  {author} {\bibinfo {author} {\bibfnamefont {S.}~\bibnamefont
  {Bravyi}}, \bibinfo {author} {\bibfnamefont {M.~B.}\ \bibnamefont
  {Hastings}}, \ and\ \bibinfo {author} {\bibfnamefont {F.}~\bibnamefont
  {Verstraete}},\ }\href@noop {} {\bibfield  {journal} {\bibinfo  {journal}
  {Phys. Rev. Lett.}\ }\textbf {\bibinfo {volume} {97}},\ \bibinfo {pages}
  {050401} (\bibinfo {year} {2006})}\BibitemShut {NoStop}%
\bibitem [{\citenamefont {Nachtergaele}\ and\ \citenamefont
  {Sims}(2010)}]{nachtergaele10}%
  \BibitemOpen
  \bibfield  {author} {\bibinfo {author} {\bibfnamefont {B.}~\bibnamefont
  {Nachtergaele}}\ and\ \bibinfo {author} {\bibfnamefont {R.}~\bibnamefont
  {Sims}},\ }\href@noop {} {\bibfield  {journal} {\bibinfo  {journal}
  {arXiv:1004.2086}\ } (\bibinfo {year} {2010})}\BibitemShut {NoStop}%
\bibitem [{SOM(text)}]{SOM}%
  \BibitemOpen
  \href@noop {} {} (\bibinfo {year} {\rm $\rm{S}$ee $\rm{S}$upplemental
  $\rm{M}$aterial at link $\rm{xxx}$ for details omitted in the main
  text})\BibitemShut {NoStop}%
\bibitem [{Note2()}]{Note2}%
  \BibitemOpen
  \bibinfo {note} {As $\alpha $ gets larger, larger separations are required
  for the algebraic decay to dominate over the exponential decay. Strictly
  speaking, this consideration imposes that Eq.\protect \tmspace +\thinmuskip
  {.1667em}\protect \textup {\hbox {\mathsurround \z@ \protect \normalfont
  (\ignorespaces \ref {eq:final_bound}\unskip \@@italiccorr )}} is only valid
  whenever $vt>\alpha \protect \qopname \relax o{log}\alpha $. This restriction
  is irrelevant, since we are ultimately concerned with the asymptotic
  light-cone shape at large $r$ and $t$, however the $\alpha \rightarrow \infty
  $ limit could in principle be taken at finite $r$ and $t$ by using the
  techniques of Ref. \cite {gong14}.}\BibitemShut {Stop}%
\bibitem [{\citenamefont {Gong}\ \emph
  {et~al.}(2014{\natexlab{b}})\citenamefont {Gong}, \citenamefont {Foss-Feig},
  \citenamefont {Clark},\ and\ \citenamefont {Gorshkov}}]{Gong_inprep_14}%
  \BibitemOpen
  \bibfield  {author} {\bibinfo {author} {\bibfnamefont {Z.-X.}\ \bibnamefont
  {Gong}}, \bibinfo {author} {\bibfnamefont {M.}~\bibnamefont {Foss-Feig}},
  \bibinfo {author} {\bibfnamefont {C.~W.}\ \bibnamefont {Clark}}, \ and\
  \bibinfo {author} {\bibfnamefont {A.~V.}\ \bibnamefont {Gorshkov}},\
  }\href@noop {} {\bibfield  {journal} {\bibinfo  {journal} {(in preparation)}\
  } (\bibinfo {year} {2014}{\natexlab{b}})}\BibitemShut {NoStop}%
\bibitem [{\citenamefont {Schachenmayer}\ \emph {et~al.}(2013)\citenamefont
  {Schachenmayer}, \citenamefont {Lanyon}, \citenamefont {Roos},\ and\
  \citenamefont {Daley}}]{schachenmayer13}%
  \BibitemOpen
  \bibfield  {author} {\bibinfo {author} {\bibfnamefont {J.}~\bibnamefont
  {Schachenmayer}}, \bibinfo {author} {\bibfnamefont {B.~P.}\ \bibnamefont
  {Lanyon}}, \bibinfo {author} {\bibfnamefont {C.~F.}\ \bibnamefont {Roos}}, \
  and\ \bibinfo {author} {\bibfnamefont {A.~J.}\ \bibnamefont {Daley}},\
  }\href@noop {} {\bibfield  {journal} {\bibinfo  {journal} {Phys. Rev. X}\
  }\textbf {\bibinfo {volume} {3}},\ \bibinfo {pages} {031015} (\bibinfo {year}
  {2013})}\BibitemShut {NoStop}%
\bibitem [{\citenamefont {Vodola}\ \emph {et~al.}(2014)\citenamefont {Vodola},
  \citenamefont {Lepori}, \citenamefont {Ercoles}, \citenamefont {Gorshkov},\
  and\ \citenamefont {Pupillo}}]{pupillo14}%
  \BibitemOpen
  \bibfield  {author} {\bibinfo {author} {\bibfnamefont {D.}~\bibnamefont
  {Vodola}}, \bibinfo {author} {\bibfnamefont {L.}~\bibnamefont {Lepori}},
  \bibinfo {author} {\bibfnamefont {E.}~\bibnamefont {Ercoles}}, \bibinfo
  {author} {\bibfnamefont {A.~V.}\ \bibnamefont {Gorshkov}}, \ and\ \bibinfo
  {author} {\bibfnamefont {G.}~\bibnamefont {Pupillo}},\ }\href@noop {}
  {\bibfield  {journal} {\bibinfo  {journal} {arXiv:1405.5440}\ } (\bibinfo
  {year} {2014})}\BibitemShut {NoStop}%
\end{thebibliography}

\end{widetext}

\end{document}